\documentclass{basi}

\usepackage{txfonts}
\usepackage{graphicx}

\def\be{\begin{equation}} \def\ee{\end{equation}}
\def\bea{\begin{eqnarray}} \def\eea{\end{eqnarray}}
 \def\be{\begin{equation}}
\def\ee{\end{equation}} \def\bea{\begin{eqnarray}}
\def\eea{\end{eqnarray}}

\def\muk{\mu{\rm K}}

\def\GeV{\,{\rm GeV}}
\def\Ghz{\,{\rm GHz}}

\def\Gpc{\,{\rm Gpc}}

\def\cmm2{{\,\rm cm^{-2}}}
\def\cm2{{\,{\rm cm}^2}}
\def\cmm3{{\,{\rm cm}^{-3}}}
\def\gcmm3{{\,{\rm g\,cm^{-3}}}}

\def\lsim{\mathrel{\rlap{\lower4pt\hbox{\hskip1pt$\sim$}}
    \raise1pt\hbox{$<$}}}                
\def\gsim{\mathrel{\rlap{\lower4pt\hbox{\hskip1pt$\sim$}}
    \raise1pt\hbox{$>$}}}                

\def\fun#1#2{\lower3.6pt\vbox{\baselineskip0pt\lineskip.9pt
  \ialign{$\mathsurround=0pt#1\hfil##\hfil$\crcr#2\crcr\sim\crcr}}}

\newcommand{\CTT}{C^{\rm TT}_\ell}
\newcommand{\CTE}{C^{\rm TE}_\ell}
\newcommand{\CEE}{C^{\rm EE}_\ell}
\newcommand{\CBB}{C^{\rm BB}_\ell}
\newcommand{\CTB}{C^{\rm TB}_\ell}
\newcommand{\CEB}{C^{\rm EB}_\ell}
\newcommand{\EI}{{\cal E}_{\rm Inf}}

\newcommand{\mpc}{\ensuremath{{\rm\,Mpc}}}

\newcommand{\Mpc}{\mpc}

\def\muk{\mu{\rm K}}

\def\biposh#1#2#3{ A^{#2}_{#3}}

\begin{document}

\title[CMB polarization]{Early Universe with  CMB polarization}
\author[T. Souradeep]%
       {Tarun Souradeep\thanks{e-mail:tarun@iucaa.ernet.in}\\
        IUCAA, Post Bag 4 , Ganeshkhind, Pune, India}

\pubyear{2011}
\volume{39}
\pagerange{\pageref{firstpage}--\pageref{lastpage}}
\date{Received 2011 April 14; accepted 2011 April 15}

\maketitle

\label{firstpage}

\begin{abstract}
The Universe is the grandest conceivable scale on which the human mind
can strive to understand nature. The amazing aspect of cosmology, the
branch of science that attempts to understand the origin and evolution
of the Universe, is that it is largely comprehensible by applying the
same basic laws of physics that we use for other branches of physics.
The observed cosmic microwave background (CMB) is understood by
applying the basic laws of radiative processes and transfer,
masterfully covered in the classic text by S. Chandrasekhar, in the
cosmological context. In addition to the now widely acclaimed temperature
anisotropy, there is also linear polarization information imprinted on
the observed Cosmic Microwave background.  CMB polarization already
has addressed, and promises to do a lot more, to unravel the deepest
fundamental queries about physics operating close to the origin of the
Universe.
\end{abstract}


\section{Introduction}\label{Intrduction}
\label{intro}

It is an honour to write an invited article commemorating the birth
centenary of Nobel laureate, Professor Subrahmanyan Chandrasekhar.
The Universe is the grandest conceivable scale on which the human mind
can strive to understand nature. Remarkably, even the origin and
evolution of the Universe is largely comprehensible by applying the
same basic laws of physics that are used in many other branches of
physics.  Chandrasekhar's research epitomizes this amazing reality,
that one can understand complex phenomena in astrophysics by building
theories based on the basic laws of physics. This article is devoted to
the cosmic microwave background (CMB), in particular, the measured
intensity and polarization fluctuations. The physics of this emerging
champion among cosmological observables is based on straightforward
application of the theory of radiative transfer of the relic radiation
from big bang through the cosmic eons -- a subject that has been
masterfully enshrined in the classic text `Radiative transfer' of
S. Chandrasekhar (1960). This text is, in fact, cited
in the seminal papers on CMB anisotropy and polarization and,
subsequent reviews (Peebles \& Yu 1970; Bond \& Efstathiou 1984, 1987; 
Bond 1996). 

Historically, theoretical development always preceded observations in
cosmology up until the past couple of decades. However, in sharp
contrast, recent developments in cosmology have been largely driven by
huge improvements in quality, quantity and the scope of cosmological
observations.  There are two distinct aspects to modern day cosmology
-- the background Universe and the perturbed Universe. The `standard'
model of cosmology must not only explain the dynamics of the homogeneous
background Universe, but also satisfactorily describe the perturbed
Universe -- the generation, evolution and finally, the formation of
the large-scale structure (LSS) in the Universe
observed in the vast galaxy surveys.  It is fair to say that cosmology
over the past few decades has increasingly seen intense interplay
between the theory and observations of the perturbed
Universe. Spectacular breakthroughs in various observations have now
concretely verified that the present edifice of the standard
cosmological models is robust.  A set of foundations and pillars of
cosmology have emerged and are each supported by a number of distinct
observations, which are listed below.

\begin{itemize}

\item{} Homogeneous, isotropic Universe, expanding from a hot initial
phase due to gravitational dynamics described by the Friedman equations 
derived from laws of General Relativity.

\item{} The basics constituents of the Universe are baryons, photons,
neutrinos, dark matter and dark energy (cosmological constant/vacuum
energy).

\item{} The homogeneous spatial sections of space-time are nearly
geometrically flat (Euclidean space).
 
\item{} Evolution of density perturbations under gravitational
instability has produced the large-scale structure in the distribution
of matter starting from the primordial perturbations in the early
Universe.

\item{} It has been established that the primordial perturbations have
correlation on length scales larger than the causal horizon; this makes a
strong case for an epoch of inflation in the very early Universe. The
nature of primordial perturbations match that expected from the
generation of primordial perturbations in simplest models of inflation.

\end{itemize}

The cosmic microwave background, a nearly uniform, thermal black-body
distribution of photons throughout space, at a temperature of 2.7
degrees Kelvin, accounts for almost the entire radiation energy
density in the Universe. Tiny variations of temperature and linear
polarization of these black-body photons of the cosmic microwave
background arriving from different directions in the sky faithfully
encode information about the early Universe. Further these photons
have travelled unimpeded across the entire observable Universe making
them excellent probes of the Universe on the largest observable
scales. The much talked about `dawn of precision era of cosmology' has
been ushered in by the study of the perturbed Universe. Measurements
of CMB anisotropy and polarization have been by far the most
influential of the cosmological observations driving advances in
current cosmology in this direction.

\section{CMB anisotropy and polarization}

\begin{figure}
\centering\includegraphics[scale=0.5,angle=-90]{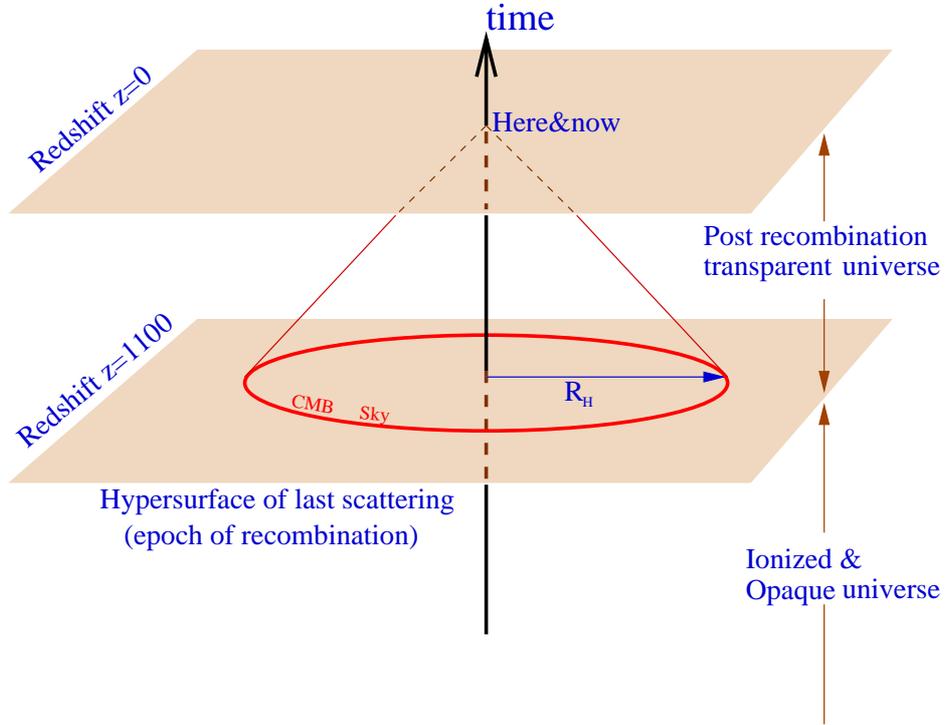}
\caption{A cartoon explaining the Cosmic Microwave Background (CMB)
using a space-(conformal) time diagram. The present Universe is
transparent and CMB photons travel to us freely over cosmic distances
along our past light cone.  In an expanding Universe, the temperature
of the Planck black-body CMB is inversely proportional to the
expansion factor. When the Universe is about $1100$ times smaller, the
CMB photons are just hot enough to keep the baryonic matter in the
Universe (about 3 quarters Hydrogen, 1 quarter Helium as determined by
big bang nucleosynthesis) ionized, and at that epoch there is a sharp
transition to an opaque Universe in the past. The CMB photons come to
us unimpeded directly from this spherical opaque surface of last
scattering at a distance of $R_H=14~\Gpc$ that surrounds us -- a super
IMAX cosmic screen. The red circle depicts the sphere of last
scattering in the reduced $2+1$ dimensional representation of the
Universe. }
\label{cmbexplain}
\end{figure}

The CMB photons arriving from different directions in the sky show
tiny variations in temperature, at a level of ten parts per million,
i.e., tens of micro-Kelvin, referred to as the CMB anisotropy, and a
net linear polarization pattern at micro-Kelvin to tens of nano-Kelvin
level.  The tiny variations of temperature and linear polarization of
these black-body photons of the cosmic microwave background arriving
from different directions in the sky have freely propagated over
cosmological distances and carry information of the the early
Universe.  As illustrated in the cartoon in Figure~\ref{cmbexplain},
the cosmic microwave background radiation sky is essentially a {\em
giant, cosmic `super' IMAX theater screen} surrounding us at a
distance of $14$ billion parsecs displaying a snapshot of the Universe
at a time very close to its origin. Hence the CMB anisotropy and
polarization are imprints of the perturbed Universe in the radiation
when the Universe was only $0.3$ millions years old, compared to its
present age of about $14$ billion years.

It is convenient to express the sky map of CMB temperature anisotropy,
$\Delta T({\bf \hat n})$ (and polarization, as we shall discuss later)
in the direction ${\bf \hat n}$ in  a spherical harmonic expansion :

\begin{equation}
\Delta T({\bf\hat n}) = \sum^\infty_{\ell = 2} \sum^\ell_{m = -\ell}
a_{\ell m} Y_{\ell m} ({\bf\hat n})\,.
\label{alm}
\end{equation} 
Theory predicts that the primary CMB anisotropy is a statistically
isotropic, Gaussian field (of zero mean), and current observations
remain fully consistent with this expectation.  The anisotropy can
then be characterized solely in terms an angular power spectrum

\begin{equation}
C_\ell= \frac{1}{(2\ell+1)}\sum^\ell_{m =
-\ell} |a_{\ell m}|^2\,.
\label{Cl}
\end{equation}

\begin{figure}[h]
\begin{center}
\includegraphics[height=6.0cm,width=8.0cm]{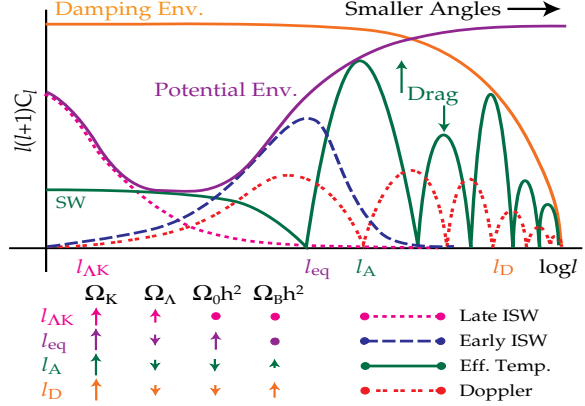}
\caption{Figure taken from Hu, Sugiyama \& Silk (1997) summarizes
    the different contributions to the primary CMB anisotropy. Also
    indicated is the dependence of the four length scales that are
    imprinted on the $C_\ell$ spectrum on some of the cosmological
    parameters.  The Sachs-Wolfe (SW) plateau at low $\ell$ is a
    faithful reproduction of the near scale-invariant spectrum of
    initial metric perturbations. Integrated Sachs-Wolfe (ISW) effect
    arises from the evolution of metric perturbations along the path
    of free streaming CMB photons.  Late ISW arises on
    $\ell<\ell_{\Lambda K}$ if the Universe has significant curvature
    or cosmological constant. The early ISW contribution at
    $\ell\sim\ell_{\rm eq}$ is due to transition from radiation to
    matter domination.  The acoustic and Doppler terms give rise to a
    harmonic series of oscillatory peaks as a snapshot of the
    oscillations of a viscous baryon-photon fluid prior to the epoch
    of recombination.  The sound horizon at recombination sets the
    length scale of the acoustic oscillations.  This `standard ruler'
    at $z\approx 1100$ then allows an accurate determination of the
    geometry of the Universe from the location of the first peak,
    $\ell_A$ via the angle-distance relationship. High baryon density
    increases viscous drag leading to suppression of even numbered
    acoustic peaks relative to odd. Power is exponentially damped at
    large $\ell$ due to photon diffusion out of matter over-densities
    (Silk damping) and finite thickness of the last scattering surface.}
\end{center}
\label{figCl}
\end{figure}

The $C_\ell$ spectra for a wide range of parameters within the
`standard' cosmology share a generic set of features neatly related to
basics physics, governing the CMB photon distribution function. On the
large angular scales (low multipole , $\ell$), the CMB anisotropy
directly probes the primordial power spectrum of metric fluctuations
(scalar gravitational potential and tensor gravitational waves) on
scales enormously larger than the `causal horizon'. On smaller angular
scales ($150<\ell<1500$), the CMB temperature fluctuations probe the
physics of the coupled baryon-photon fluid through the imprint of the
acoustic oscillations in the ionized plasma sourced by the same
primordial fluctuations. At even higher multipoles, the damping tail
of the oscillations encodes interesting physics such as the slippage
in the baryon-photon coupling, temporal width of the opaque to
transparent Universe transition, weak lensing due to large scale
structures in the Universe.  Fig.~\ref{figCl} that dissects the CMB
angular power spectrum attempts to provide a cryptic summary of the
various kinds of physics involved. Overall, the physics of CMB
anisotropy has been very well understood for more than two decades,
Furthermore, the predictions of the primary anisotropy and linear
polarization and their connection to observables are, by and large,
unambiguous (Bond 1996; Hu \& Dodelson 2002).

The acoustic peaks occur because the cosmological perturbations excite
acoustic waves in the relativistic plasma in the early Universe.  The
recombination of baryons at redshift $z\approx 1100$ effectively
decouples the baryon and photons in the plasma abruptly switching off
the wave propagation.  In the time between the excitation of the
perturbations and the epoch of recombination, modes of different
wavelength can complete different numbers of oscillation periods, or
in other words, waves can travel a finite distance and then
freeze. This translates the characteristic time scale into a
characteristic length scale and leads to a harmonic series of maxima
and minima in the CMB anisotropy power spectrum.  The acoustic
oscillations have a characteristic scale known as the sound horizon,
which is the comoving distance that a sound wave could have traveled
up to the epoch of recombination. This well-determined physical scale
of $150 \Mpc$ is imprinted on the CMB fluctuations at the surface of
last scattering, the typical scale of the random bright and dull patches
on the `cosmic super-IMAX' screen.

\begin{figure}
\centering\includegraphics[scale=0.4]{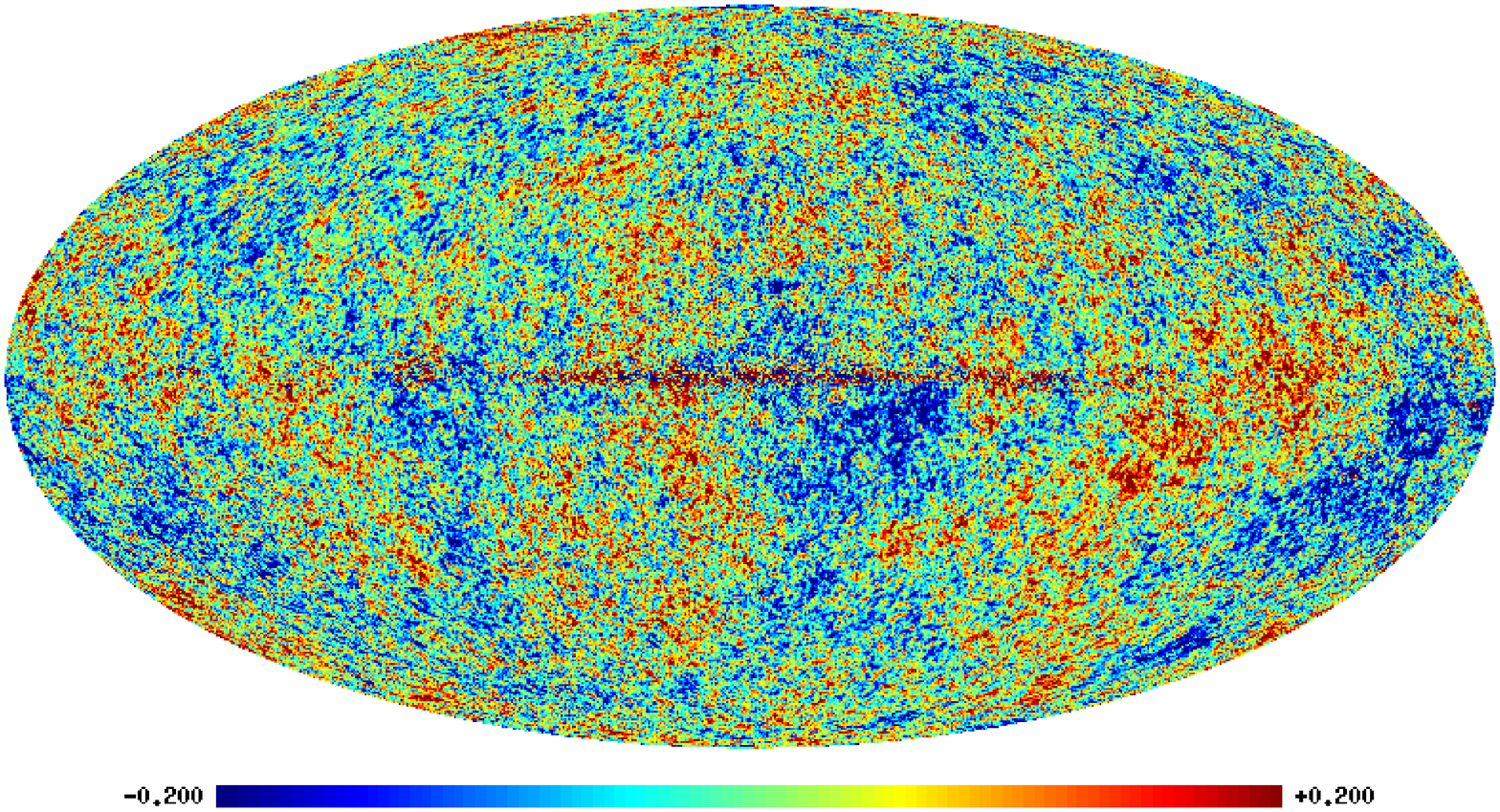}
\centering\includegraphics[scale=0.4]{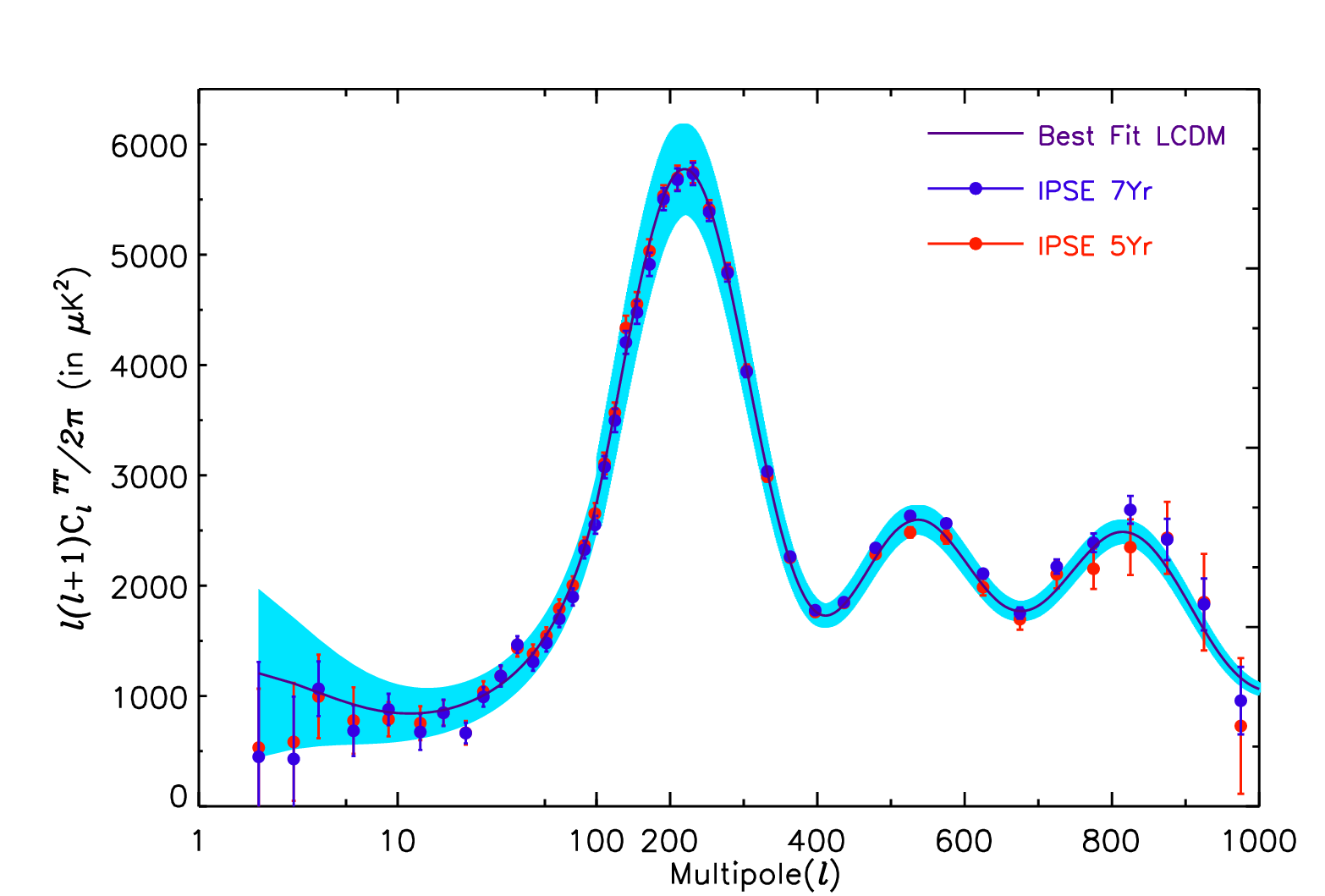}
\caption{\footnotesize The exquisite temperature anisotropy data from the three
years of data from the WMAP satellite is shown in the figures.  {\em Top:}
The top figure shows colour-coded full sky map (in Mollewide
projection) of the CMB temperature variations. The temperature
variations range between $\pm 200\muk$ with an rms of
about $70\muk$. The angular resolution of features of the map is about
a quarter of a degree. For comparison, the first CMB anisotropy
measurements in 1992 by the DMR instrument on board the COBE satellite
produced the same map at a much coarser resolution of $7$ degrees.
{\em Bottom:} The angular power spectrum estimated from the
multi-frequency five- and seven-year WMAP data. The result from IPSE, a
self-contained model free approach to foreground removal
(Saha, Jain \& Souradeep 2006; Samal et al. 2010)  matches that obtained by the WMAP
team. The solid curve showing prediction of the best fit power-law,
flat, $\Lambda$CDM model threads the data points
closely [Figure: courtesy Tuhin Ghosh]. }
\label{WMAP13}
\end{figure}

The angle subtended by this physical scale in the CMB sky (IMAX
screen) at a known distance of $14\Gpc$ then allows a sensitive
determination of the geometry ($\Omega_{0K}$) of the background
Universe. Essentially, the same standard ruler of $150\Mpc$ placed at
$14 \Gpc$ would subtend different angles in a Universe with different
spatial curvature.  This determines the location of the series of
harmonic peaks of $C_\ell$ along the multipole $\ell$ seen in
Fig.~\ref{figCl}.  The amplitude of baryon-photon oscillations can be
expected to directly scale with the amount of baryons available in the
Universe.  Consequently, the height of the peaks in the $C_\ell$
sensitively determine the baryon density, $\Omega_B$.  The $C_\ell$s are
sensitive to other important cosmological parameters, such as, the
relative density of matter, $\Omega_m$, cosmological constant,
$\Omega_\Lambda$, Hubble constant, $H_0$ and deviation from flatness
(curvature), $\Omega_K$.  Implicit in $C_\ell$ is the hypothesized
nature of random primordial/initial metric perturbations -- (Gaussian)
statistics , (nearly scale-invariant) power spectrum, (largely)
adiabatic vs. iso-curvature and (largely) scalar vs. tensor
component.  The `default' settings in bracket are motivated by
inflation (Starobinsky 1982; Guth \& Pi 1982; Bardeen, Steinhardt \&
Turner 1983).

Besides, the entirely theoretical motivation of the paradigm of
inflation, the assumption of Gaussian, random adiabatic scalar
perturbations with a nearly scale-invariant power spectrum is arguably
also the simplest possible theoretical choice for the initial
perturbations.  What has been truly remarkable is the extent to which
recent cosmological observations have been consistent with and, in
certain cases, even vindicated the simplest set of assumptions for the
initial conditions for the perturbed Universe discussed below.

The first two decades ($\sim$~1991-2011) of exciting CMB anisotropy
measurements have been capped off with the release of 7 years of data
from the Wilkinson Microwave Anisotropy Probe (WMAP) of
NASA~\footnote{Wilkinson Microwave Anisotropy Probe mission
http://wmap.gsfc.nasa.gov/}. The first detection of CMB anisotropy by
COBE-DMR in 1992 observationally established the origin and mechanism
of structure formation in the Universe. Observations were then made at
three frequencies, $90$, $53$ and $31~\Ghz$ which allowed a fairly
good removal of the `foreground' contamination of the cosmic signal by
the strong emission from our own Galaxy. The $15$-years old
experimental success story of CMB anisotropy measurements, starting
from discovery of CMB anisotropy by the COBE satellite in 1992, has
been topped off by the exquisite data from the WMAP.  The WMAP
satellite was placed at the second Lagrange point of the Sun-Earth
system.  Measurements from WMAP combine high angular resolution with
full sky coverage and high sensitivity due to the stable thermal
environment allowed by a space mission. Moreover, observations were
made at five frequencies, $94$ (W-band), $61$ (V-band), $41$ (Q-band),
$33$ (Ka-band) and $23~\Ghz$ (K-band) that allowed much better
removal of the `foreground' contamination. Similar to the
observational strategy of COBE-DMR, the satellite measures CMB
temperature differences between a pair of points in the sky. Each day
the satellite covered $30\%$ of the sky, but covers the full sky in
$6$ months. This massive redundancy in measurements allows the mission
to beat down the detector noise to from milli-Kelvins to tens of
micro-Kelvin level. The WMAP mission has acquired data for about nine
years up until August 2010 and made that public at regular intervals
after a short proprietary possession (first year data was released in
2003, three year data in 2006, five year data in 2008, and seven year
data in 2010). A final data release of the entire nine years of data
in expected in the coming year.

The measured angular power spectrum of the cosmic microwave background
temperature fluctuations, $C_\ell$, shown in Fig.~\ref{WMAP13} has
become invaluable for constraining cosmological models. The position
and amplitude of the peaks and dips of the $C_\ell$ are sensitive to
important cosmological parameters.  The most robust constraint
obtained is that on the spatial curvature of the Universe and baryon
density.  The observations establish that space on cosmic scales is
geometrically flat ($\Omega_K =0$) to within sub-percent
precision. The dominant energy content in the present Universe is a
mysterious matter with negative pressure dubbed, dark energy, or, the
cosmological constant, which contributes about $73\%$ of the total
energy budget ($\Omega_\Lambda=0.73$), followed by cold non-baryonic
dark matter about $23\%$ ($\Omega_m=0.23$) and, most humbly, ordinary
matter (baryons) account for only about $4\%$ ($\Omega_B=0.04$) of the
matter budget.  The current up to date status of cosmological
parameter estimates from joint analysis of CMB anisotropy and
large-scale structure (LSS) data is usually found in the parameter
estimation paper accompanying the most recent results of a major
experiment, such as the recent WMAP release of 7-year data (Komatsu et
al. 2011; Larson et al. 2011).

More recently, CMB polarization measurements have provided the
required complementary information on the nature of initial conditions
for the primordial fluctuations.  One of the firm predictions of the
working `standard' cosmological model is a random pattern of linear
polarization ($Q$ and $U$ Stokes parameters) imprinted on the CMB at
last scattering surface. Thomson scattering generates CMB polarization
anisotropy at decoupling (Bond \& Efstathiou 1984; Hu \& White 1997).
This arises from the polarization dependence of the differential cross
section: $d\sigma/d\Omega\propto |\epsilon'\cdot\epsilon|^2$, where
$\epsilon$ and $\epsilon'$ are the incoming and outgoing polarization
states involving linear polarization only (Rybicki \& Lightman 1979).
A local quadrupole temperature anisotropy produces a net polarization,
because of the $\cos^2\theta$ dependence of the cross section.  A net
pattern of linear polarization is retained due to local quadrupole
intensity anisotropy of the CMB radiation impinging on the electrons
at the last scattering surface. The polarization pattern on the sky
can be decomposed in the two kinds with different parities.  The even
parity pattern arises as the gradient of a scalar field called the
$E$--mode.  The odd parity pattern arises from the `curl' of a
pseudo-scalar field called the $B$--mode of polarization.  The
observed CMB sky map is then characterized by a triplet of random
scalar fields: $X(\hat{n})\equiv \{\Delta T(\hat{n})$, $E(\hat{n})$,
$B(\hat{n})\}$.  It is possible to generalize equation (\ref{alm}) to
express both CMB anisotropy and polarization in spherical harmonic
space as

\begin{equation}
X({\bf\hat n}) = \sum^\infty_{\ell = 2} \sum^\ell_{m = -\ell}
a^X_{\ell m} Y_{\ell m} ({\bf\hat n})\,.
\label{almX}
\end{equation} 
and also define a set of observable angular power spectra analogous to
eqn.~(\ref{Cl}) as

\begin{equation}
C^{XX'}_\ell= \frac{1}{(2\ell+1)}\sum^\ell_{m = -\ell} a^X_{\ell
  m}a^{X'*}_{\ell m}\,.
\label{ClX}
\end{equation} 
For statistically isotropic, Gaussian CMB sky, there are a total of 4
power spectra that characterize the CMB signal~: $\CTT, \CTE, \CEE,
\CBB$. Parity conservation within standard radiative processes
eliminates the two other possible power spectra, $\CTB$ \& $\CEB$.
Important point to note is that the odd-parity B-mode of polarization
cannot sourced by scalar density perturbations, or potential velocity
flow. B mode polarization can arise only due to shear fields acting on
photon distribution, such as, from gravitational waves and (weak)
gravitational lensing deflection of photons.

After the first detection of CMB polarization spectrum by the Degree
Angular Scale Interferometer (DASI) on the intermediate band of
angular scales ($l\sim 200-440$) in late 2002 (Kovac et al. 2002), the
field has rapidly grown, with measurements coming in from a host of
ground--based and balloon--borne dedicated CMB polarization
experiments.  The full sky E-mode polarization maps and polarization
spectra from WMAP were a new milestone in CMB
research (Kogut et al. 2003; Page et al. 2007).  Although the CMB polarization
is a clean probe of the early Universe that promises to complement the
remarkable successes of CMB anisotropy measurements it is also a much
subtler signal than the anisotropy signal. Measurements of
polarization by ongoing experiments at sensitivities of $\mu K$
(E-mode) have had to overcome numerous challenges in the past
decade. The tens of $nK$ level B-mode signal pose the ultimate
experimental and analysis challenge to this area of observational
cosmology.  The most current CMB polarization measurement of $\CTT$,
$\CTE$ and $\CEE$ and a non--detection of $B$--modes come from QUaD
and BICEP. They also report interesting upper limits $\CTB$ or $\CEB$,
over and above observational artifacts (Wu et al. 2009). A non-zero
detection of $\CTB$ or $\CEB$, over and above observational artifacts,
could be tell-tale signatures of exotic parity violating
physics (Lue, Wang \& Kamionkowski 1999; Maity, Mazumdar \& Sengupta 2004)
  and the CMB measurements put interesting limits
on these possibilities.

The immense dividends of CMB polarization measurements for
understanding the physics behind the origin and evolution of our
Universe have just started coming in.  While CMB temperature
anisotropy can also be generated during the propagation of the
radiation from the last scattering surface, the CMB polarization
signal can be generated primarily at the last scattering surface,
where the optical depth of the Universe transits from large to small
values. The polarization information complements the CMB temperature
anisotropy by isolating the effect at the last scattering surface from
other distinct physical effects acting during the propagation of the
photons along the line of sight.

The polarization measurements provide an important test on the
adiabatic nature of primordial scalar fluctuations~\footnote{Another
independent observational test comes from the recent measurements of
the Baryon Acoustic Oscillations (BAO) in the power spectrum of LSS in
the distribution of galaxies. BAO has also observationally established
the gravitational instability mechanism for structure formation.}.  CMB
polarization is sourced by the anisotropy of the CMB at recombination,
consequently, the angular power spectra of temperature and
polarization are closely linked. The power in the CMB polarization
signal is sourced by the gradient (velocity) term in the same acoustic
oscillations of the baryon-photon fluid at last scattering that gives
rise to temperature (intensity) anisotropy. Hence, a clear evidence of
adiabatic initial conditions for primordial fluctuations is that the
compression and rarefaction peaks in the temperature anisotropy
spectrum should be `out of phase' with the gradient (velocity) driven
peaks in the polarization spectra.

\begin{figure}[h]
\begin{center}
  \includegraphics[scale=0.45, angle=-90]{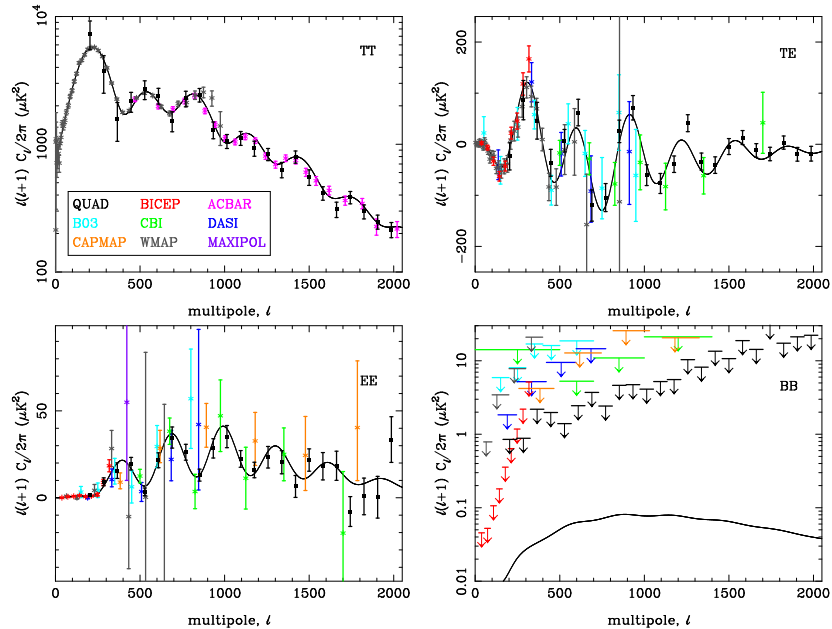}
\caption{Figure taken from Brown et al. (2009)  shows a
compilation of recent measurements of the angular power spectra of CMB
anisotropy and polarization from a number of CMB experiments.  The
data is good enough to indicate that the peaks in EE and TE are out of
phase with that of TT as expected for adiabatic initial
conditions. The null BB detection of primary CMB signal from gravity
waves is not unexpected (given the ratio of tensor to scalar
perturbations expected in the simplest models of inflation). It is
important to note that the upper limits on $\CBB$ have improved by
almost an order of magnitude in the past two years from experiments
such as QUAD and BICEP. }
\label{cmbspecpol}
\end{center}
\end{figure}

Figure~\ref{cmbspecpol} taken from Brown et al. (2009) reflects the
current observational status of CMB E-mode polarization
measurements. The recent measurements of the angular power spectrum
the E-mode of CMB polarization at large $l$ have confirmed that the
peaks in the $\CEE$ spectra are out of phase with that of the
temperature anisotropy spectrum $\CTT$.

While the power in the CMB temperature anisotropy at low multipoles
($l\lsim 60$) first measured by the COBE-DMR (Smoot et al. 1992) did
point to the existence of correlated cosmological perturbations on
super Hubble-radius scales at the epoch of last scattering, it left
open the (rather unlikely) `logical' alternative possibility that all
the power at low multipole is generated through integrated Sachs-Wolfe
effect along the line of sight later in the Universe (when the Hubble
scale is larger). However, since the polarization anisotropy is
generated only at the last scattering surface, the negative trough
clearly visible at high significance in the $\CTE$ spectrum at $l\sim
130$ (that corresponds to a scale larger than the horizon at the epoch
of last scattering) sealed this loophole, and provides an unambiguous
proof of apparently `acausal' correlations in the cosmological
perturbations. This was first first measured by WMAP and later
reconfirmed with higher significance by QUaD and BICEP (Kogut et
al. 2003; Bennett et al. 2003; Page et al. 2007; Brown et al. 2009;
Chiang et al. 2010).

The B-mode CMB polarization is a very clean and direct probe of the
early Universe physics that generated the primordial metric
perturbations.  Inflationary models necessarily produce tensor
perturbations (gravitational waves) that are predicted to evolve
independently of the scalar density perturbations, with an
uncorrelated power spectrum.  The tensor modes on the scales of
Hubble-radius the line of sight to the last scattering distort the
photon propagation and generate an additional anisotropy pattern
predominantly on the largest scales.  (The amplitude of a tensor mode
falls off rapidly on sub-Hubble radius scales, hence is important on
angular scales comparable to larger than the Hubble radius at last
scattering). It is common to parametrize the tensor component by the
ratio $r_{k_*} = A_{\rm t}/A_{\rm s}$, ratio of $A_{\rm t}$, the
primordial power in the transverse traceless part of the metric tensor
perturbations, and $A_{\rm s}$, the amplitude scalar perturbation at a
comoving wave-number, $k_*$ (in $\mpc^{-1}$).  For power-law models,
recent WMAP data alone puts an improved upper limit on the tensor to
scalar ratio, \ensuremath{r_{0.002} < 0.55 \mbox{ } (95\%\mbox{\ CL})}
and the combination of WMAP and the lensing-normalized SDSS galaxy
survey implies \ensuremath{r_{0.002} < 0.28 \mbox{ } (95\%\mbox{\
CL})} (MacTavish et al. 2006).

\begin{figure}[h]
\begin{center}
\includegraphics[scale=0.5]{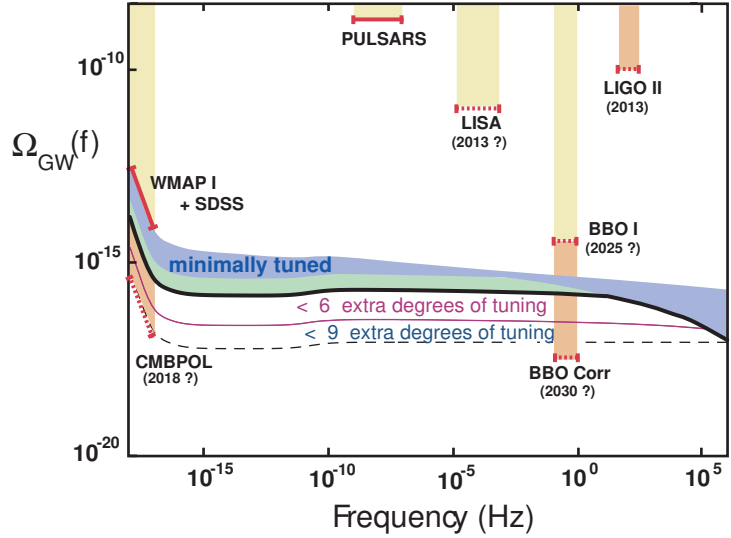} 
\caption{The figure taken from Boyle, Steinhardt \& Turok 2006, and
NASA/DOE/NSF Task force report on Cosmic Microwave
Background research, 2005 (http://www.nsf.gov/mps/ast/tfcr.jsp)  
shows the theoretical predictions and observational constraints on
primordial gravitational waves from inflation. The gravitational wave
energy density per logarithmic frequency interval, (in units of the
critical density) is plotted versus frequency. The shaded (blue) band
labeled `minimally tuned' represents the range predicted for simple
inflation models with the minimal number of parameters and
tunings. The dashed curves have lower values of tensor contribution,
$r$, that is possible with more fine tuned inflationary scenarios.
The currently existing experimental constraints shown are due to: big
bang nucleosynthesis (BBN), binary pulsars, and WMAP-1 (first year)
with SDSS. Also shown are the projections for LIGO (both LIGO-I, after
one year running, and LIGO-II); LISA; and BBO (both initial
sensitivity, BBO-I, and after cross-correlating receivers,
BBO-Corr). Also seen the projected sensitivity of a future space
mission for CMB polarization (CMBPol).}
\label{SGWBspec} 
\end{center}
\end{figure}

On angular scales corresponding to the multipole range $50<\ell<150$,
the B (curl) component of CMB polarization is a unique signature of
tensor perturbations from inflation.  The amplitude of tensor
perturbation is directly proportional to Hubble parameter during
inflation, Loosely speaking, this is related to the Hawking temperature
in de-Sitter like space-times. In turn, $H_{\rm inf}$ is related to the
energy density $\EI$ of the Universe during inflation through the
Friedman equation governing cosmological evolution. Hence, the CMB
B-polarization is a direct probe of the energy scale of early Universe
physics that generated the primordial metric perturbations (scalar \&
tensor). The relative amplitude of tensor to scalar perturbations,
$r$, sets the energy scale for inflation $\EI = 3.4\times
10^{16}$~\GeV~$r^{1/4}$.  A measurement of $B$--mode polarization on
large scales would give us this amplitude, and hence {\em a direct
determination of the energy scale of inflation.}  Besides being a
generic prediction of inflation, the cosmological gravity wave
background from inflation would be also be {\em a fundamental test of
GR on cosmic scales and the semi--classical behavior of gravity}.
Figure~\ref{SGWBspec} summarizes the current theoretical
understanding, observational constraints, and future possibilities for
the stochastic gravity wave background from inflation. The stochastic
gravitational wave background from inflation is expected to exist on
cosmological scales down to terrestrial scales. The first CMB
normalized GW spectra from inflation using the COBE results was given
by Souradeep \& Sahni (1992) from IUCAA.  This prediction
will be targeted by both CMB polarization experiments, as well as,
future GW observatories in space, such as Big Bang Observatory (BBO),
DECIGO and LISA (Marx et al. 2010).

Gravitational lensing of the stronger CMB E-polarization by ongoing
process of structures formation along the line of sight to the last
scattering surface also generates a significant $B$--mode polarization,
but on smaller angular scales($\ell > 200$). This is prediction is
shown as the black curve in bottom left panel for $\CBB$ measurements
in Fig.~\ref{cmbspecpol}. The lensing signal carries important
information about the matter power spectrum and its evolution over a
range of redshift inaccessible to other observations.  This promises a
powerful probe for constraining the nature of dark energy and, also
more excitingly, for determining the neutrino masses. Recent studies
indicate that measuring the lensing polarization signal to the cosmic
variance limit, can potentially place limits on the total mass of
neutrinos at a level comparable to the measured mass differences from
neutrinos oscillations.

While there has been no detection of cosmological signal in B-mode of
polarization, the lack of $B$--mode power suggests that foreground
contamination from polarized emission from our own Galaxy is at a
manageable level and is very encouraging news for the prospects of
future measurements.  The Planck satellite launched in May 2009 will
greatly advance our knowledge of CMB polarization by providing
foreground/cosmic variance--limited measurements of $\CTE$ and $\CEE$
out beyond $l\sim 1000$.  We also expect to detect the weak lensing
signal in $\CBB$, although with relatively low precision, that
required for placing ultimate limits on the total neutrino
mass. Perhaps, Planck could also detect the stochastic inflationary
gravitational wave background if it exists at a level of $r\sim
0.1$. Dedicated future CMB polarization space missions are under study
at both NASA and ESA in the time frame 2020+~\footnote{ NASA/DOE/NSF
Task force report on Cosmic Microwave Background research,
2005. http://www.nsf.gov/mps/ast/tfcr.jsp (Also available at the
Legacy Archive for Microwave Background Data analysis (LAMBDA) site
http://lambda.gsfc.nasa.gov).}.  Lower budget missions would primarily
target the low multipole $B$-mode polarization signature of gravity
waves and consequently, identify the viable sectors in the space of
inflationary parameters. More ambitious plans such proposal, such as
COrE~\footnote{Cosmic Origins Explorer (COrE) proposal,
http://www.core-mission.org}, target the entire useful $\CBB$ spectrum
and also plan to probe other exciting results from CMB weak lensing
measurements.

\section{Beyond the angular power spectra of the CMB sky}
  
It is well appreciated that in `classical' big bang model the initial
perturbations would have had to be generated `acausally'. Besides
resolving a number of other problems of classical Big Bang, inflation
provides a mechanism for generating these apparently `acausally'
correlated primordial perturbations (Starobinsky 1982; Guth \& Pi
1982; Bardeen et al. 1983). There is increasing effort towards
establishing this observationally. There are subtle observations of
the CMB sky that could reveal more clearly the mechanism for
generations of primordial fluctuation, or, perhaps surprise us by
producing insurmountable challenges to the inflation paradigm.

\subsection{Statistical isotropy of the Universe}
\label{SI}

The {\em Cosmological Principle} that led to the idealized FRW
Universe found its strongest support in the discovery of the (nearly)
isotropic, Planckian, cosmic microwave background. The isotropy around
every observer leads to spatially homogeneous cosmological models.
The large scale structure in the distribution of matter in the
Universe (LSS) implies that the symmetries incorporated in FRW
cosmological models ought to be interpreted statistically. These are
also predicted in the simplest models of inflation.

The CMB anisotropy and its polarization is currently the most
promising observational probe of the global spatial structure of the
Universe on length scales close to, and even somewhat beyond, the
`horizon' scale ($\sim c H_0^{-1}$).  The exquisite measurement of the
temperature fluctuations in the CMB provide an excellent test bed for
establishing the statistical isotropy (SI) and homogeneity of the
Universe. In `standard' cosmology, CMB anisotropy signal is expected
to be statistically isotropic, i.e., statistical expectation values of
the temperature fluctuations $\Delta T(\hat q)$ are preserved under
rotations of the sky. In particular, the angular correlation function
$C(\hat{q},\, \hat{q}^\prime)\equiv\langle\Delta T(\hat q)\Delta
T(\hat q^\prime)\rangle$ is rotationally invariant for Gaussian
fields. In spherical harmonic space, where $\Delta T(\hat q)=
\sum_{lm}a_{lm} Y_{lm}(\hat q)$, the condition of {\em statistical
isotropy} (SI) translates to a diagonal $\langle a_{lm} a^*_{l^\prime
m^\prime}\rangle=C_{l} \delta_{ll^\prime}\delta_{mm^\prime}$ where
$C_l$, is the widely used angular power spectrum of CMB anisotropy.
The $C_l$ is a complete description only of (Gaussian) SI CMB sky CMB
anisotropy and  would be (in principle) an inadequate measure for
comparing models when SI is violated (Bond, Pogosyan \& Souradeep
1998, 2000a,b).

Interestingly enough, the statistical isotropy of CMB has come under a
lot of scrutiny after the WMAP results. Tantalizing evidence of SI
breakdown (albeit, in very different guises) has mounted in the {\it
WMAP} first year sky maps, using a variety of different statistics. It
was pointed out that the suppression of power in the quadrupole and
octupole are aligned (Tegmark, de Oliveira-Costa \& Hamilton 2004).
Further ``multipole-vector'' directions associated with these
multipoles (and some other low multipoles as well) appear to be
anomalously correlated (Copi, Huterer \& Starkman 2004; Schwartz et
al. 2004).  There are indications of asymmetry in the power spectrum
at low multipoles in opposite hemispheres (Eriksen et al. 2004).
Analysis of the distribution of extrema in {\it WMAP} sky maps has
indicated non-Gaussianity, and to some extent, violation of SI (Larson
\& Wandelt 2004).  The more recent WMAP maps are consistent with the
first-year maps up to a small quadrupole difference. The additional
years of data and the improvements in analysis has not significantly
altered the low multipole structures in the maps (Hinshaw et
al. 2007). Hence, `anomalies' persisted at the same modest level of
significance and are unlikely to be artifacts of noise, systematics,
or the analysis in the first year data.  The cosmic significance of
these `anomalies' remains debatable also because of the aposteriori
statistics employed to ferret them out of the data. The WMAP team has
devoted an entire publication to discuss and present a detailed
analysis of the various anomalies (Bennett et al. 2011).

The observed CMB sky is a single realization of the underlying
correlation, hence detection of SI violation, or correlation patterns,
pose a great observational challenge. It is essential to develop a
well defined, mathematical language to quantify SI and the ability to
ascribe statistical significance to the anomalies unambiguously.  The
Bipolar spherical harmonic (BipoSH) representation of CMB correlations
has proved to be a promising avenue to characterize and quantify
violation of statistical isotropy.

Two point correlations of CMB anisotropy, $C(\hat{n}_1,\, \hat{n}_2)$,
are functions on $S^2 \times S^2$, and hence can be generally expanded
as \be
\label{bipolar} C(\hat{n}_1,\, \hat{n}_2)\, =\, \sum_{l_1,l_2,\ell,M}
\biposh{}{\ell M}{l_1 l_2} Y^{l_1l_2}_{\ell M}(\hat{n}_1,\,
\hat{n}_2)\,. \ee Here $\biposh{}{\ell M}{l_1 l_2}$ are the Bipolar
Spherical harmonic (BipoSH) coefficients of the expansion and
$Y^{l_1l_2}_{\ell M}(\hat{n}_1,\, \hat{n}_2)$ are bipolar spherical
harmonics. Bipolar spherical harmonics form an orthonormal basis on
$S^2 \times S^2$ and transform in the same manner as the spherical
harmonic function with $\ell,\, M$ with respect to
rotations. Consequently, inverse-transform of $C(\hat{n}_1,\,
\hat{n}_2)$ in eq.~(\ref{bipolar}) to obtain the BipoSH coefficients
of expansion is unambiguous.

Most importantly, the Bipolar Spherical Harmonic (BipoSH)
coefficients, $\biposh{}{\ell M}{l_1 l_2}$, are linear combinations of
{\em off-diagonal elements} of the harmonic space covariance matrix,
\be
\label{ALMvsalm} \biposh{}{\ell M}{l_1 l_2} \,=\, \sum_{m_1m_2}
\langle a_{l_1m_1}a^{*}_{l_2 m_2}\rangle (-1)^{m_2} {\mathcal C}^{\ell
M}_{l_1m_1l_2 -m_2} \ee where ${\mathcal C}_{l_1m_1l_2m_2}^{\ell M}$
are Clebsch-Gordan coefficients and completely represent the information of
the covariance matrix.  

Statistical isotropy implies that the covariance matrix is diagonal, $
\langle a_{lm}a^{*}_{l' m'}\rangle = C_{l}\,\, \delta_{ll^\prime}
\delta_{mm'}$ and hence the angular power spectra carry all
information of the field. When statistical isotropy holds BipoSH
coefficients, $\biposh{}{\ell M}{ll'}$, are zero except those with
$\ell=0, M=0$ which are equal to the angular power spectra up to a
$(-1)^l (2l+1)^{1/2}$ factor.  Therefore to test a CMB map for
statistical isotropy, one should compute the BipoSH coefficients for
the maps and look for nonzero BipoSH coefficients. {\em Statistically
significant deviations of BipoSH coefficient of map from zero would
establish violation of statistical isotropy.}

Since $A^{\ell M}_{l_1 l_2}$ form an equivalent representation of a
general two point correlation function, cosmic variance precludes
measurement of every individual $A^{\ell M}_{l_1 l_2}$. There are
several ways of combining BipoSH coefficients into different
observable quantities that serve to highlight different aspects of SI
violations.  Among the several possible combinations of BipoSH
coefficients, the Bipolar Power Spectrum (BiPS) has proved to be a
useful tool with interesting features (Hajian \& Souradeep 2003;
Hajian, Souradeep \& Cornish 2005). BiPS of CMB
anisotropy is defined as a convenient contraction of the BipoSH
coefficients \be
\label{kappal} \kappa_\ell \,=\, \sum_{l,l',M} W_l W_{l'}\left|\biposh{}{\ell
M}{ll'}\right|^2 \geq 0 \ee where $W_l$ is the window function that
corresponds to smoothing the map in real space by symmetric kernel to
target specific regions of the multipole space and isolate the SI
violation on corresponding angular scales.

The BipoSH coefficients can be summed over $l$ and $l'$ to reduce the
cosmic variance, to as obtain reduced BipoSH (rBipoSH)
coefficients (Hajian \& Souradeep 2006) \be \label{first} A_{\ell M}=
\sum_{l=0}^{\infty}\sum_{l'=|\ell-l|}^{\ell+l} \biposh{}{\ell
M}{ll'}. \ee Reduced bipolar coefficients orientation information of
the correlation patterns. An interesting way of visualizing these
coefficients is to make a {\em Bipolar map} from $A_{\ell M}$ \be
\Theta(\hat{n}) = \sum_{\ell=0}^{\infty}\sum_{M=-\ell}^{\ell} A_{\ell
M} Y_{\ell M} (\hat{n}).  \ee The symmetry $A_{\ell M}=(-1)^M A_{\ell
-M}^*$ of reduced bipolar coefficients guarantees reality of
$\Theta(\hat{n})$.

\begin{figure}[h]
\begin{center}
\includegraphics[scale=0.7]{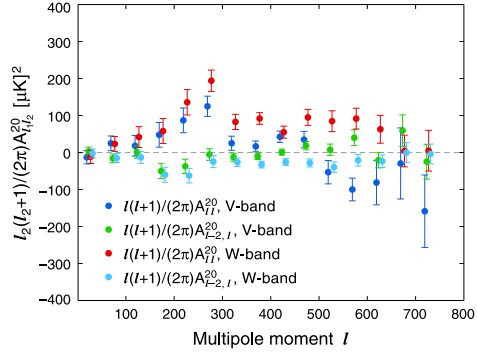}
\caption{Figure taken from WMAP-7 yr publication on anomalies in the
CMB sky (Bennett et al. 2011) shows the measured quadrupolar (bipolar
index $L=2$) bipolar power spectra for V-band and W-band WMAP data,
using the KQ75y7 mask. The spherical multipole have been binned within
uniform bands $\delta l = 50$. Only the components of the bipolar
power spectra with M = 0 in ecliptic coordinates are shown. A statistically
significant quadrupolar effect is seen, even for a single frequency
band in a single angular bin.}
\end{center}
\label{WMAP7_biposh} 
\end{figure}

It is also possible to obtain a measurable band power measure of
$A^{\ell M}_{l_1 l_2}$ coefficient by averaging $l_1$ in bands in
multipole space. Recently, the WMAP team has chosen to quantify SI
violation in the CMB anisotropy maps by the estimation $A^{\ell M}_{l
l-i}$ for small value of bipolar multipole, $L$, band averaged in
multipole $l$. Fig.~\ref{WMAP7_biposh} taken from the WMAP-7 release
paper (Bennett et al. 2011) shows SI violation measured in WMAP CMB maps

CMB polarization maps over large areas of the sky have been recently
delivered by experiments in the near future. The statistical isotropy
of the CMB polarization maps will be an independent probe of the
cosmological principle.  Since CMB polarization is generated at the
surface of last scattering, violations of statistical isotropy are
pristine cosmic signatures and more difficult to attribute to the
local Universe.  The Bipolar Power spectrum has been defined and
implemented for CMB polarization and show great promise (Basak, Hajian
\& Souradeep 2006; Souradeep, Hajian \& Basak 2006).

\subsection{Gaussian primordial perturbations}
\label{gauss}

The detection of primordial non-Gaussian fluctuations in the CMB would
have a profound impact on our understanding of the physics of the
early Universe. The Gaussianity of the CMB anisotropy on large angular
scales directly implies Gaussian primordial
perturbations
(Munshi, Souradeep \& Starobinsky 1995; Spergel \& Goldberg 1999)
that is theoretically motivated by
inflation. The simplest inflationary models predict
only very mild non-Gaussianity that should be undetectable in the
WMAP data.

The CMB anisotropy maps (including the non Gaussianity analysis
carried out by the WMAP team data; Komatsu et al. 2011) have been found
to be consistent with a Gaussian random field.  Consistent with the
predictions of simple inflationary theories, no significant deviations
from Gaussianity in the CMB maps using general tests such as Minkowski
functionals, the bispectrum, trispectrum in the three year WMAP
data
(Spergel et al. 2007; Komatsu et al. 2011).
There have however been numerous
claims of anomalies in specific forms of non-Gaussian signals in the
CMB data from WMAP at large scales (see discussion in sec.~\ref{SI}).
Recently, a new class of odd-parity bispectra has been discovered
enriching the field significantly (Kamionkowski \& Souradeep 2011).

\section{Summary}

The past few years has seen the emergence of a `concordant'
cosmological model that is consistent both with observational
constraints from the background evolution of the Universe as well that
from the formation of large scale structures.  It is certainly fair to
say that the present edifice of the `standard' cosmological models is
robust. A set of foundation and pillars of cosmology have emerged and
are each supported by a number of distinct
observations (Ostriker \& Souradeep 2004; Souradeep 2011).

 Besides precise determination of various parameters of the `standard'
cosmological model, observations have also established some important
basic tenets of cosmology and structure formation in the Universe --
`acausally' correlated initial perturbations, adiabatic nature
primordial density perturbations, and gravitational instability as the
mechanism for structure formation. The favored, concordance model
inferred is a spatially flat accelerating Universe where structures
have formed by the gravitational evolution of nearly scale invariant,
adiabatic perturbations, as expected from inflation.  The signature of
primordial perturbations observed through the CMB anisotropy and
polarization is the most compelling evidence for new, possibly
fundamental, physics in the early Universe that underlie the scenario
of inflation (or related alternatives). Searches are also on for
subtle signals in the CMB maps beyond the angular power spectrum in
the violation of statistical isotropy
(Hajian \& Souradeep 2003, 2006; Hajian, Souradeep \& Cornish 2005; 
Basak, Hajian \& Souradeep 2006; Souradeep, Hajian \& Basak 2006),
or, in the violation of Gaussianity
(Munshi, Souradeep \& Starobinksy 1995; Spergel \& Goldberg 1999).

Cosmology is a branch of physics that has seen theoretical enterprise
at its best. During the long period of sparse observations in its
history, brilliant theoretical ideas (and prejudices) shaped a
plausible self-consistent scenario. But current cosmology is passing
through a revolution.  In the recent past, cosmology has emerged as a
data rich field increasingly driven by exquisite and grand
observations of unprecedented quality and quantity. These observations
have transformed cosmology into an emergent precision science of this
century. Further, CMB polarization is arguably emerging as a key
observable that can also address fundamental questions related to the
origin of the Universe.

\section*{Acknowledgments}

I would like to thank the editors, D. J. Saikia and Virginia Trimble
for inviting me to write this article and for their immense patience
in obtaining it from me. It is indeed a privilege to contribute to a
commemorative volume honoring an iconic scientist,
Prof. S. Chandrasekhar, whom I had the good fortune to meet, and talk
to, early in my research career at IUCAA. Finally, I acknowledge my
past and present students and collaborators who have contributed to my
own research in this exciting area.


\end{document}